\documentclass[prd,nofootinbib,preprint]{revtex4}
\usepackage[T1]{fontenc}
\usepackage{amsmath,amssymb}
\usepackage{epsfig}
\graphicspath{{Figs/}}
\usepackage{dcolumn}
\usepackage{bm}
\usepackage{graphicx}
\usepackage{subcaption}
\usepackage[usenames,dvipsnames]{color}
\usepackage{adjustbox}
\usepackage{slashed}
\usepackage[colorlinks,citecolor=blue]{hyperref}
\usepackage{pdfpages}
\usepackage{float}
\usepackage{multirow}
\usepackage[autostyle]{csquotes}

\begin{document}
\title{Exploring the feasibility of  the charged lepton flavor violating decay $ \mu \rightarrow e + \gamma $ in inverse and linear seesaw mechanisms with $A_4$ flavour symmetry}
	
	\author{Maibam Ricky Devi}
	\email{deviricky@gmail.com}
	
	\author{Kalpana Bora}
	\email{kalpana@gauhati.ac.in}
	\affiliation{Department of Physics, Gauhati University, Guwahati-781014, Assam, India}
	
\begin{abstract}
One of the possible ways to explain the observed flavour structure of fundamental particles is to include flavor symmetries in the theories. In this work, we investigate the rare charged lepton flavour violating (cLFV) decay process ($\mu \rightarrow e\gamma$) in two of the low scale ($\sim$TeV)  seesaw models: (i) the Inverse seesaw (ISS) and (ii) Linear seesaw (LSS) models within the framework of $A_{4}$ flavour symmetry.  Apart from the $ A_{4} $ flavour symmetry, some other symmetries like $U(1)_{X}$, $Z_4$ and $Z_5$ are included to construct the Lagrangian. We use results from our previous work \cite{Devi:2021ujp,Devi:2021aaz} where we computed unknown neutrino oscillation parameters within $3 \sigma$ limits of their global best fit values, and apply those results to compute the branching ratio (BR) of the muon decay for both the seesaw models. Next we compare our results  with the current experimental bounds and sensitivity limits of BR($\mu \rightarrow e\gamma$) as projected by various experiments, and present a comparative analysis that which of the two models is more likely to be tested by which current/future experiment. This is done for various values of currently allowed non-unitarity parameter. This comparative study will help us to pinpoint that which of the low scale seesaw models and triplet flavon VEV alignments will be more viable and favourable for testing under a common flavour symmetry ($A_{4}$ here), and hence can help discriminate between the two models. 
\end{abstract}
	\maketitle

\section{Introduction}
\label{sec:1}
The hunt for the charged lepton flavour violating (cLFV) decay $ \mu \rightarrow e +\gamma $ is like finding a needle in a stack of hay and has been a challenging endeavour for particle physicists. Lepton family number is not conserved in transitions among families in these processes. In standard model (SM), lepton flavour is conserved at all orders of perturbation theory, but may get nonzero contributions through some beyond SM processes, like neutrino oscillation. The MEG (Mu to E Gamma) experiment \cite{Baldini:2013ke, MEG:2016leq} (2008-2013) at the Paul Scherrer Institute (PSI), located at Villigen, Switzerland has set the most stringent upper limit on $BR( \mu \rightarrow e + \gamma) \leq 4.2 \times 10^{-13} $ at $ 90 \% $ C.L. in its first phase. This constraint would either validate or nullify models that predict the golden channel of cLFV decay, $ \mu \rightarrow e +\gamma $  by incorporating physics beyond the Standard Model. The future sensitivity of this channel is expected to get enhanced up to an order of $ BR( \mu \rightarrow e + \gamma) < 5.0 \times 10^{-14} $ by the upgraded version of MEG experiment, also popularly called MEG II experiment \cite{Cattaneo:2017psr} which is currently in commissioning phase. The rare muon decay channel like $ \mu \rightarrow e+\gamma $ is called \enquote{golden channel} - firstly due to the abundant rate of muon production in cosmic radiation as well as in accelerators and secondly because of its significantly longer lifetime than the rest of the leptons \cite{Kettle:2013lwa}. It is interesting to note that the cLFV processes can be related to neutrino oscillation, due to the mixing of neutrino flavour eigenstates within loop diagrams. It is known that though such cLFV decay rates are highly suppressed in SM, however get enhanced if there is mixing between light and heavy neutrino eigenstates. This linkage of neutrino mixing to cLFV processes can directly give us an unambiguous signature of new physics, and therefore has been of interest to the scientific community. Although the manifestation of new physics is quite unclear, its detection can open up a new portal to our understanding of baryon asymmetry of the universe, flavour structure of neutrino mixing etc.\\
\\
Recently, an intense positive and negative muon beam facility powered by Proton Improvement Plan II, or PIP-II \cite{CGroup:2022tli} is created in the Fermilab accelerator complex. The Advanced Muon Facility (AMF) \cite{CGroup:2022tli} will drive cutting-edge research on studies related to cLFV to answer some of its most profound questions. The HiMB (High Intensity Muno Beam)\cite{Aiba:2021bxe} could further increase the sensitivity of  MEG II experiment once it's ready for phase one.  Hence theoretical studies related to $\mu \rightarrow e+\gamma$ become even more interesting and relevant.\\
\\
Keeping in view the above discussion on the importance of $\mu \rightarrow e+\gamma$ decays, in this work we compute and analyse the $ BR(\mu \rightarrow e + \gamma)$ for different VEV alignments of $ A_{4} $ triplet flavons, for linear (LSS) and inverse low-scale seesaw (ISS) models. Since high-scale seesaw models are not directly accessible to experiments, we choose low-scale seesaw models\cite{Devi:2021ujp, Devi:2021aaz}. Currently not much is known about symmetry breaking scale and VEV alignment etc. of $A_{4} $ flavour symmetry, also it is not clear which seesaw mechanism is more favourable. It is possible to obtain an observable cLFV decay rate satisfying the current experimental bounds and future sensitivity through neutrino oscillation, thus we can explore the feasibility of the two low-scale seesaw models for the detection of the cLFV decay under a common $ A_{4} $ flavour symmetry. Some other similar works can be found in \cite{Deppisch:2002vz, Mandal:2022zmy,  Deppisch:2004xv, Hesketh:2022wgw, MuonCollider:2022xlm, Bu:2014boa, Pascoli:2016wlt, Heinrich:2018nip}. Through this work, one can also comment on which of the two models would be more favourable in context of BR of cLFV $\mu \rightarrow e+\gamma$ decay. 
\\
\\
We proposed an inverse seesaw model in the Refs\cite{Devi:2021ujp, Devi:2021aaz}, where we have studied the correlation between effective neutrino mass of $ 0\nu\beta \beta $ decay and $ m_{lightest} $ by using the unknown neutrino oscillation data, i.e., $ m_{lightest} $, $ \delta_{CP} $, $ \alpha $ and $ \beta $ obtained from our model and their corresponding known  neutrino oscillation parameters taken from recent global fit data \cite{deSalas:2020pgw}. Using these parameter data we have pinpointed the favored octant and mass hierarchy for the favored VEV alignment of the triplet scalar flavon involved in our model. In this work we have successfully pinpointed that for (-1,1,1)/(1,-1,-1) VEV alignment of the   triplet scalar flavon, the correlation between $ m_{lightest} $ and effective neutrino mass of $ 0\nu \beta \beta $ aligns with the correlation plot of  $ m_{lightest} $ and $ m_{0\nu \beta \beta} $ obtained from experimental global analysis. We also constructed the linear seesaw model in \cite{Devi:2021ujp} where we have compared our previously mentioned  ISS model and our linear model. One can find  other $ A_{4} $ flavour symmetry based neutrino models \cite{Dinh:2016tu,Chen:2012st, Kalita:2015jaa, Sarma:2018bgf}. In \cite{Hirsch:2009mx, Sruthilaya:2017mzt}, an inverse as well as linear  models are presented with detailed discussion of symmetries such as $SU(2)_L $, $Z_{3}$ and $ A_{4} $ for model building. Many other seesaw based neutrino models are presented in 
\cite{Borah:2017dm, Sahu:2020tqe} where $ A_{4} $ symmetry is exclusively used to construct neutrino model in a specific seesaw scenario.\\
\\
To find the unknown neutrino oscillation parameters, i.e., $ m_{lightest} $, $ \delta_{CP} $, $ \alpha $ and $ \beta $, we have compared the light neutrino mass matrix obtained from each model with the light neutrino mass matrix, $ m_{\nu}= U^{T}_{PMNS}m_{\nu diag} U_{PMNS}$ where $ U_{PMNS} $ is the PMNS mixing matrix. After comparison of these matrices, we get a set of equations for both NH and IH for 26 possible combinations of VEV alignment of the triplet scalar flavon (TSF) involved in the models. We then solve these equations simultaneously  to find the unknown parameters. A parameter scan of the rest of the known neutrino parameters, i.e., mixing angles, squared mass differences, is done within the $ 3\sigma $ range of global fit data. We choose the solutions which are very precise by checking them with a tolerance of $ <10^{-5} $. This precision allow us to predict only a few favored VEV alignments in the TSF with specific mass hierarchy and octant of the atmospheric mixing angle $ \theta_{23} $. \\
\\
To maintain the appropriate symmetries, the flavon VEVs must align in a specific manner which can be derived from the minimization of the full scalar potential.
The so-called F-term alignment mechanism (as mentioned in Ref. \cite{Altarelli:2005yx}) is the most widely used technique for producing unique flavon VEVs. In a supersymmetric configuration, the flavons are intended to be coupled to so-called driving fields. Driving fields, like flavons, transform generally in a non-trivial fashion under the family symmetry G while remaining neutral under the SM gauge group. The F-term equations are frequently solved for the trivial vacuum, which is the vacuum configuration in which none of the flavons produce a VEV. By incorporating soft supersymmetry breaking effects, this can be avoided and it is possible to get more or less any VEV alignment. \\
\\
In our work we have chosen the possible cases of VEV alignments through the minimization of the scalar potential. The minimization of the full scalar potential and its possible VEV alignments for both ISS and LSS models is shown in Appendices  A.1 and A.2. For all the flavons in our setup, the most generic renormalizable potential that is invariant under $ A_{4}\times Z_{4} \times Z_{5} \times U(1)_{X} $ in case of ISS model and  $ A_{4}\times Z_{5} \times Z^{\prime}_{5} $  for LSS model have been stated which on minimization with respect to the different components of the triplet scalar field $ \Phi_{s} $, gives us a set of equations. We can determine the possible VEV alignments of the triplet scalar flavon field by solving those set of equations. The allowed and unallowed cases of  these VEV alignments of TSF is shown in Table (\ref{vevalignmentcases}). The details of this are available in our earlier work \cite{Devi:2021aaz}. \\
\\
This paper has been arranged as follows. In section \ref{sec:2} we present  a brief introduction to charged lepton flavour violation $ \mu \rightarrow e + \gamma $ decay. The ISS and LSS models that will be used are constructed and presented in details in section \ref{sec:3}. In section \ref{sec:num} we discuss the numerical analysis to compute branching ratio of  $ \mu \rightarrow e + \gamma $ decay using the unknown neutrino oscillation parameters  as computed in \cite{Devi:2021aaz} in both the models. In section \ref{sec:result}  we present the results  and a discussion on them. Section \ref{sec: concl} contains summary and conclusions.


\section{cLFV decay ($\mu \rightarrow e + \gamma $) }
\label{sec:2}
The charged lepton flavour violating transitions could provide us with a direct signature of new physics beyond the Standard Model. Search for such lepton flavour violating decay processes have been studied in a host of channels in ongoing experiments such as MEG collaboration at PSI. However, no cLFV decay processes have been detected so far, as their decay rates are highly suppressed. Out of various decay channels, the most sensitive transitions are the ones involving first and second generation of leptons especially muons, i.e., $ \mu^{+} \rightarrow e^{+} + \gamma $, $ \mu^{-} N \rightarrow e^{-} N $, $ \mu^{+} \rightarrow e^{+}e^{-}e^{+}$ because of their abundance in cosmic radiation and particle accelerators \cite{Calibbi:2017uvl}. The flavour structure of neutrino mass matrix can help understand the charged lepton flavour violation too. The decay rate of  $ \mu \rightarrow e + \gamma $ can be expressed as \cite{Bilenky:1977du, LalAwasthi:2011aa, Parida:2016asc, He:2002pva, DelleRose:2015bms, Forero:2011pc,  MarcanoImaz:2017xjc, Dolan:2018qpy,  Cheng:2000ct} :
\begin{equation}
BR(l_{\alpha} \rightarrow l_{\beta} \gamma ) \approx \dfrac{\alpha^{3}_{W} sin^{2}\theta_{W} m^{5}_{l\alpha }}{256 \pi^{2} M^{4}_{W} \Gamma_{l \alpha} }  \left|  \sum\limits^{9}_{i=1} \mathcal{K}^{*}_{\alpha i}\mathcal{K}_{\beta i} G\left( \dfrac{\mathcal{M}^{2}_{i}}{M^{2}_{W}} \right) \right|^{2},
\label{BR form}
\end{equation}
Here, the parameter $\alpha_W=g^2/4\pi$, and $g$ represents  weak coupling. Also, $\theta_W$ is the electroweak mixing angle, $M_W$ is the mass of $W^{\pm}$ boson, $ \mathcal{M}_{i} $ is the mass of neutrinos (both light and heavy neutrinos), $ m_{l\alpha } $ is the mass of the decaying charged lepton $l_{\alpha}$ and finally, $\Gamma_{l_{\alpha}}$ is the total decay width of the decaying charged lepton $l_{\alpha}$. $ \mathcal{K}_{(\alpha,\beta) i} $ ($ \alpha =\mu, \beta = e $ with i=1,..,9) define the elements of the matrix $ \mathcal{K}$ that block  diagonalises the $ 9\times 9 $ neutrino mass matrix $ M_{\nu} $ \cite{Dolan:2018qpy} of the inverse seesaw and linear seesaw models, and further analysis is shown later in sub-sections \ref{subsec:ISS}  and \ref{subsec:LSS}  respectively. The form factor $G(x)$ is \cite{MarcanoImaz:2017xjc, Dolan:2018qpy, Forero:2011pc, Cheng:2000ct, Ilakovac:1994kj, Alonso:2012ji} :
\begin{equation}
   G(x)= -\dfrac{2x^{3}+5x^{2}-x}{4(1-x)^{3}}-\dfrac{3x^{3}lnx}{2(1-x)^{4}} \textrm{, where }  x=\dfrac{\mathcal{M}^{2}_{i}}{M^{2}_{W}}
\end{equation}
The matrix $\mathcal{K}$ can be expressed as \cite{Dolan:2018qpy} :
\begin{equation}
\begin{aligned}
 \mathcal{K} =  \left(  \begin{matrix}
\mathbb{1}-\dfrac{1}{2}B^{*}B^{T} &   B^{*}\\
-B^{T} & \mathbb{1}-\dfrac{1}{2}B^{T}B^{*}
\end{matrix} \right)\left(  \begin{matrix}
U &   0\\
0 & V
\end{matrix} \right)
\end{aligned}
\label{eqn:K}
\end{equation}
where U is the usual PMNS matrix that block diagonalises the light neutrinos and V is the unitary matrix that diagonalises the heavy right-handed neutrinos \cite{Dolan:2018qpy}. The full parametrisation of the $ 9\times 9 $ active sterile flavour mixing matrix can be found in \cite{Han:2021qum}.

As will be seen in the inverse seesaw model in Eqn. (\ref{ISSmnu}) in the following sections, we can repartition the $ 9\times 9 $ matrix into a type-I like matrix as
\begin{equation}
(M_{\nu})_{iss}=\left( \begin{matrix}
0 & M_{D} \\
M^{T}_{D} & M_{R}
\end{matrix} \right) 
\end{equation}
where
\begin{equation}
\begin{aligned}
M_{D} & =\left( \begin{matrix}
m_{D} & 0
\end{matrix} \right) \textrm{, }
M_{R}= \left( \begin{matrix}
0 & M \\
M^{T} & \mu_{s}
\end{matrix} \right) \textrm{, }
M^{-1}_{R}= \left( \begin{matrix}
-X^{-1} & M^{T-1} \\
M^{-1} & 0
\end{matrix} \right) \\
\textrm{ where, } & X=M\mu^{-1}_{s}M^{T}.
\end{aligned}
\label{eqn:1}
\end{equation}
Similarly in the case of linear seesaw model, we can repartition Eqn. (\ref{LSS: LSS mass matrix}) as
\begin{equation}
(M_{\nu})_{lss}=\left( \begin{matrix}
0 & M^{\prime}_{D} \\
M^{\prime T}_{D} & M^{\prime}_{R}
\end{matrix} \right) 
\end{equation}
where,
\begin{equation}
\begin{aligned}
 M^{\prime}_{D}= \left(  \begin{matrix}
M_{D} &   M_{L}
\end{matrix} \right) \textrm{ and } M^{\prime}_{R}= \left(  \begin{matrix}
0 &   M\\
M^{T} & 0
\end{matrix} \right) \\ \Rightarrow 
 M^{\prime -1}_{R}  =  \left(  \begin{matrix}
0 &   M^{T-1}\\
M^{-1} & 0
\end{matrix} \right) 
\end{aligned} 
\label{LSS:3}
\end{equation}
In Eqn. (\ref{eqn:K}), the entry $B$ acts a small perturbation matrix which can  expressed as \cite{Dolan:2018qpy}:\\
\begin{equation}
\begin{aligned}
& &   B= M_{D}M^{-1}_{R}=(m_{D}M^{T-1}\mu_{s} M^{-1},m_{D}M^{T-1})
 &  & \textrm{ \textit{(in case of ISS)}}
  \end{aligned}
  \label{B:ISS}
\end{equation}
\begin{equation}
\begin{aligned}
 & &  B= M^{\prime}_{D}M^{\prime -1}_{R}= (M_{L}M^{-1},M_{D}M^{T-1})  
&  &  & \textrm{ \textit{(in case of LSS)}}
\end{aligned}
  \label{B:LSS}
\end{equation} 
 The matrix V can be numerically computed as \cite{Dolan:2018qpy, Karmakar:2016cvb}
\begin{equation}
V= \dfrac{\sqrt{2}}{2}\left(\begin{matrix}
\mathbb{1}_{3} & -i\mathbb{1}_{3}\\
\mathbb{1}_{3} & i\mathbb{1}_{3}
\end{matrix}\right) + \mathcal{O}(\mu_{s} M^{-1})
\label{V}
\end{equation}
The matrix element $ \mu_{s} $ is absent in the neutrino mass matrix of linear seesaw model. The matrix $ \mathcal{K}_{3\times 3} $ is related to the parameter $ \eta $\cite{Dolan:2018qpy} that represents deviation from unitarity (for both ISS and LSS models) as\\
\begin{equation}
\mathcal{K}_{3\times 3}=(\mathbb{1} - \dfrac{1}{2}B^{*}_{i}B^{T}_{i})U =(\mathbb{1} - \eta )U
\label{K3x3}
\end{equation}
\begin{equation}
\textrm{where, } \eta= \dfrac{1}{2}B^{*}_{i}B^{T}_{i} = \dfrac{1}{2}\vert M_{D}M^{T-1}\vert^{2} 
\label{non unitarity}
\end{equation}
\begin{table}
\centering
\begin{tabular*}{\columnwidth}{@{\extracolsep{\fill}}lllll@{}}
\hline
\multicolumn{1}{@{}l}{Experiments} & Year & Upper Limit  & Ref.\\
\hline

   MEG & 2016 &  $ 4.2 \times 10^{-13} $ & \cite{Baldini:2013ke, MEG:2016leq}\\
   MEG II & Commissioned in 2017 &  $ 5.0 \times 10^{-14 } $ $ ^{*} $ & \cite{Cattaneo:2017psr} \\
AMF (PIP II in FermiLab) & 2022 &  \textit{(In planning stage)} $ ^{**} $ & \cite{CGroup:2022tli} \\
    
\hline
\end{tabular*}
\caption{\textit{Current upper limit (MEG) and future sensitivity of $ BR(\mu \rightarrow e + \gamma) $ set by various experiments. $( ^{*} )$ indicates the future sensitivity of MEG II experiment, and $ ( ^{**}) $ that of the the Advanced Muon Facility. }}
\label{tab:2}
\end{table}
\section{ The Models}
\label{sec:3}
As stated earlier, this work aims to constrain and compare the ISS and LSS models with $A_4$ flavour symmetry with reference to the cLFV decay $(\mu \rightarrow e + \gamma)$, and to check their testability and viability at ongoing/future planned experiments. To compute the branching ratio of the muon decay, we use these models with $A_{4} $ symmetry from our previous works \cite{Devi:2021ujp, Devi:2021aaz}. 
\subsection{Inverse seesaw model } 
\label{subsec:ISS}
The $ 9\times 9 $ neutrino mass matrix in the basis ($ \nu^{c}_{L}, N, S $) obtained from inverse seesaw (ISS) mechanism is \cite{Dev:2009aw}
\begin{equation}
 M_{\nu}=\begin{bmatrix}
0 & M_{D} & 0\\
M^{T}_{D} & 0 & M\\
0 & M^{T} & \mu_{s}
\end{bmatrix} \textrm{.}
\label{ISS mass matrix}
\end{equation}
We present here the ISS model, for the sake of completeness of the work, that contains a  $ SU(2)_{L}$ singlet right-handed neutrino N along with three other singlet fermions $ S_{i=1,2,3}$ (Sterile neutrinos) apart from the Standard Model particles. The particle content of this model under $ A_{4}\times Z_{4} \times Z_{5} \times U(1)_{X} $ symmetry is given in the Table \ref{tab:vevA4ISS} below \cite{Devi:2021ujp,Devi:2021aaz}.\\
\vspace{0.1 in}
\begin{table}[h]
\begin{center}
\begin{tabular}{|c|cc|ccc|cc|cccccccc|}
\hline
 & L & $\mathcal{H}$ & $ e_{R} $ & $ \mu_{R} $ & $ \tau_{R} $ & N & S & $\Phi_T$ & $\Phi_s$ & $\Omega$ & $\xi$ & $\tau$ & $\rho$ & $\rho'$ & $\rho^{\prime\prime}$\\
\hline
$A_4$ & 3 & 1 & 1 & $ 1^{\prime\prime} $ & $ 1^{\prime} $ & 3 & 3 & 3 & 3 & 1 & $1'$ & $1''$ & 1 & 1 & 1 \\
\hline 
$Z_4$ & 1 & 1 & i & i  & i & i & 1 & i & -i & -i & -i & -i & i & i & 1 \\
\hline
$Z_{5}$ & 1  & 1 & $ \omega $ & $ \omega $  & $ \omega $ & $ \omega^{2} $ & 1 & $ \omega$ & 1 & 1 & 1 &1 & $ \omega^{2} $ & 1 & 1 \\
\hline
$U(1)_{X}$ & -1 & 0 & -1 & -1 & -1 & -1 & 1 & 0 & -1 & -1 & -1 & -1 & 0 & -4 & -3 \\
\hline
\end{tabular}
\end{center}
\caption{\textit{Particle content for inverse seesaw model  under $ A_{4}\times Z_{4} \times Z_{5} \times U(1)_{X} $ symmetry  \cite{Devi:2021ujp,Devi:2021aaz}}}
\label{tab:vevA4ISS}
\end{table}
In Table (\ref{tab:vevA4ISS}) L and $\mathcal{H}$ represents the charged lepton family and the  $ SU(2)_{L} $ Higgs doublet respectively, whereas, $\Omega$,  $\xi$,  $\tau$,  $\rho$,  $\rho'$ and $\rho^{\prime\prime}$ are the $ A_{4} $ singlet flavons and $\Phi_s$  is the $ A_{4} $ triplet scalar.  We choose the vev of flavon $\phi_T$ as $\langle \phi_T \rangle=\upsilon_T(1,0,0)$ \cite{Altarelli:2005yx} so that the charged lepton mass matrix turns out to be diagonal in the leading order  as   
$M_{l} = \upsilon_{h} \frac{\upsilon^{\dagger}_{T}}{\Lambda} {\rm{diag}}\left( Y_{e},  Y_{\mu}, Y_{\tau} \right)$ where $  Y_{e}$, $ Y_{\mu}$, $Y_{\tau}  $ represent the Yukawa coupling constants. 
The relevant Lagrangian for the neutrino sector is given as:\\
\begin{equation} 
\mathcal{L}_{\rm Y} \supset  Y_D \frac{\bar{L} \tilde{\mathcal{H}} N \rho^{\dagger}}{\Lambda} + Y_M N S \rho^{\dagger} + Y_{\mu_{s}} S S [\frac{\rho^{\prime} \rho^{\prime\prime^{\dagger}}(\Phi_s + \Omega + \xi + \tau) }{\Lambda^2}  ]+ h.c. \textrm{,}
\label{ISS:revised lagrangian}
\end{equation}
\\
Using above equation, the mass matrix elements in Eqn. (\ref{ISS mass matrix}) can be written in the form as:
\begin{equation}
 M_{D}=\dfrac{Y_{D}\upsilon_{h}\upsilon^{\dagger}_{\rho}}{\Lambda}\begin{bmatrix}
1 & 0 & 0\\
0 & 0 & 1\\
0 & 1 & 0
\end{bmatrix} \textrm{, }
  M=Y_{M}\upsilon^{\dagger}_{\rho}\begin{bmatrix}
1 & 0 & 0\\
0 & 0 & 1\\
0 & 1 & 0
\end{bmatrix}  \textrm{, }\\
\label{ISS: MD}
\end{equation}
\begin{equation}
 \textrm{and, } \mu_{s}=\dfrac{Y_{\mu_{s}}\upsilon_{\rho^{\prime}}\upsilon^{\dagger}_{\rho_{\prime\prime}}}{\Lambda^{2}}\left(
\begin{array}{ccc}
v_{\Omega }+2 v_s \phi _a & v_{\xi }-v_s \phi _c & v_{\tau }-v_s \phi _b \\
 v_{\xi }-v_s \phi _c & v_{\tau }+2 v_s \phi _b & v_{\Omega }-v_s \phi _a \\
 v_{\tau }-v_s \phi _b & v_{\Omega }-v_s \phi _a & v_{\xi }+2 v_s \phi _c \\
\end{array}
\right) \textrm{.}
\label{ISS:Mu}
\end{equation}
Here, $\Lambda$ is the usual cutoff scale of the theory. $ Y_D $, $ Y_M $, $ Y_{\mu_{s}} $ are the dimensionless coupling constants which are usually complex. The non-zero VEVs of scalars can be represented as:  $ \langle \mathcal{H} \rangle= v_{h} $,     $ \langle \Omega \rangle= v_{\Omega} $, 
$ \langle\rho \rangle= v_{\rho} $, $ \langle\rho^{\prime} \rangle= v_{\rho^{\prime}} $, $ \langle\rho^{\prime \prime} \rangle= v_{\rho^{\prime \prime}} $, $ \langle \xi \rangle= v_{\xi} $,  $ \langle \tau \rangle= v_{\tau} $,  $ \langle\Phi_{S} \rangle= v_{s}(\Phi_{a},\Phi_{b},\Phi_{c}) $. The light neutrino mass matrix for inverse seesaw model is computed as \cite{Dev:2009aw} :
\begin{equation}
m_{\nu}= M_{D}(M^{T})^{-1}\mu_{s} M^{-1}M^{T}_{D} \textrm{ .}
\label{ISS:neutrino_matrix_equation}
\end{equation}
Next, using the matrix elements of Eq.(\ref{ISS:revised lagrangian}) into Eq.(\ref{ISS:neutrino_matrix_equation}),  the light neutrino mass matrix is obtained as,\\
\begin{equation}
 \Rightarrow m_{\nu}=F_{1}\left(
\begin{array}{ccc}
 v_{\Omega }+2 v_s \phi _a & v_{\xi }-v_s \phi _c & v_{\tau }-v_s \phi _b \\
 v_{\xi }-v_s \phi _c & v_{\tau }+2 v_s \phi _b & v_{\Omega }-v_s \phi _a \\
 v_{\tau }-v_s \phi _b & v_{\Omega }-v_s \phi _a & v_{\xi }+2 v_s \phi _c \\
\end{array}
\right)\textrm{,}
\label{ISS:final matrix with vev}
\end{equation}
where, $ F_{1}= \dfrac{Y^{2}_{D}Y_{\mu_{s}}}{Y^{2}_{M}}[\dfrac{v^{2}_{h}v_{\rho^{\prime}}v^{\dagger}_{\rho^{\prime \prime}}}{\Lambda^{4}}]$.
\subsection{Linear seesaw model } 
\label{subsec:LSS}
The neutrino mass matrix using linear seesaw (LSS) mechanism  \cite{Akhmedov:1995ip, Malinsky:2005bi} with the basis ($\nu_L, N_R^c, S_R^c$) is given as \\
\begin{equation}
 M_{\nu}=\begin{bmatrix}
0 & M_{D} & M_{L}\\
M^{T}_{D} & 0 & M\\
M^{T}_{L} & M^{T} & 0
\end{bmatrix} \textrm{.}
\label{LSS: LSS mass matrix}
\end{equation}\\
For the LSS model we use $ A_{4}\times Z_{5} \times Z^{\prime}_{5} $ symmetries to generate the tiny but non-zero neutrino masses. The $ A_{4} $ singlet flavons are ($ \varepsilon , \kappa ,\zeta , \varphi , \varphi^{\prime} $), and other  fields of the model under  $ A_{4}\times Z_{5} \times Z^{\prime}_{5}  $ symmetry such as $ A_{4} $ triplet scalar $\Phi_s$, $ SU(2)_{L} $ charged lepton doublets L and $ SU(2)_{L} $ Higgs doublet $ \mathcal{H} $  are shown in Table \ref{tab:vevA4LSS}. The effective light neutrino mass matrix formula for linear seesaw is given as  \cite{Deppisch:2015cua}, \\
\begin{table}
\begin{center}
\begin{tabular}{|c|cc|ccc|cc|ccccccc|}
\hline
 & L & $ \mathcal{H} $ & $ e_{R} $ & $ \mu_{R} $ &$ \tau_{R} $  & $ \mathcal{N} $ & $ \mathcal{S}$ & $\varepsilon$ & $ \Phi_{T} $ & $\Phi_s$ & $\kappa$ & $\zeta$ & $\varphi$ &  $\varphi^{\prime}$\\
\hline
$A_{4}$ & 3 & 1 & 1 & $1^{\prime\prime}  $& $ 1^{\prime} $ & 3 & 3 & 1 & 3 & 3 & 1 & $1'$ & $1''$  & 1 \\
\hline 
$Z_{5}$ & $ \omega $ & 1 & $ \omega $ & $ \omega $ & $ \omega $  & $ \omega^{2} $ & $ \omega^{3} $ &$ \omega $ & 1 & $ \omega^{2} $ & $ \omega^{2} $ &$ \omega^{2} $ & $ \omega^{2} $ & 1 \\
\hline
$Z^{\prime}_{5}$ & $ \omega $ & 1 & $ \omega $ & $ \omega $ & $ \omega $ & $ \omega $ & $ \omega^{2} $ & 1 & 1 & $ \omega $ & $ \omega $ & $ \omega $ & $ \omega $ & $ \omega^{2} $ \\
\hline
\end{tabular}
\end{center}
\caption{\textit{Particle content for  linear seesaw model under $ A_{4}\times Z_{5} \times Z^{\prime}_{5} $  symmetry }}
\label{tab:vevA4LSS}
\end{table}
\begin{equation}
m_{\nu}= M_{D}(M_{L}M^{-1})^{T}+(M_{L}M^{-1})M^{T}_{D}\textrm{.}
\label{Mu_matrix_equation}
\end{equation}
Taking into consideration the transformation of the neutrino fields under  $ A_{4}\times Z_{5} \times Z^{\prime}_{5} $  symmetry and its interaction with the other fields, the Lagrangian of the neutrino sector can be now be written as:\\
\begin{equation} 
\mathcal{L}_{\nu} \supset  Y_D \frac{\bar{L} \tilde{\mathcal{H}} N \varepsilon^{\dagger}}{\Lambda} + Y_M N S\varphi^{\prime}+ Y_{L} \frac{\bar{L} \tilde{\mathcal{H}} S}{\Lambda} (\Phi_s^{\dagger} + \kappa^{\dagger} + \zeta^{\dagger} + \varphi^{\dagger}) \textrm{.}
\label{LSS:lagrangian modified}
\end{equation}
The mass matrix elements in Eqn. (\ref{LSS: LSS mass matrix}) become:\\
\begin{equation}
 M_{D}=\dfrac{Y_{D}\upsilon_{h}\upsilon^{\dagger}_{\varepsilon}}{\Lambda}\begin{bmatrix}
1 & 0 & 0\\
0 & 0 & 1\\
0 & 1 & 0
\end{bmatrix} \textrm{, }
M=Y_{M}\upsilon_{\varphi^{\prime}}\begin{bmatrix}
1 & 0 & 0\\
0 & 0 & 1\\
0 & 1 & 0
\end{bmatrix}\textrm{,}
\label{linearmassmatrix1}
\end{equation}
\begin{equation}
  M_{L}=\dfrac{Y_{L}\upsilon_{h}}{\Lambda}\begin{bmatrix}
2\upsilon^{\dagger}_{s}\Phi_{a} + \upsilon^{\dagger}_{\kappa} & -\upsilon^{\dagger}_{s}\Phi_{c} + \upsilon^{\dagger}_{\varphi} & -\upsilon^{\dagger}_{s}\Phi_{b} + \upsilon^{\dagger}_{\zeta} \\
-\upsilon^{\dagger}_{s}\Phi_{c}+ \upsilon^{\dagger}_{\varphi}  & 2\upsilon^{\dagger}_{s}\Phi_{b}+ \upsilon^{\dagger}_{\zeta} & -\upsilon^{\dagger}_{s}\Phi_{a}+ \upsilon^{\dagger}_{\kappa}   \\
-\upsilon^{\dagger}_{s}\Phi_{b}+ \upsilon^{\dagger}_{\zeta} & -\upsilon^{\dagger}_{s}\Phi_{a}+ \upsilon^{\dagger}_{\kappa} & 2\upsilon^{\dagger}_{s}\Phi_{c} +  \upsilon^{\dagger}_{\varphi} 
\end{bmatrix}\textrm{.}
\label{linearmassmatrix2}
\end{equation}
The matrix elements from Eqn. (\ref{LSS:lagrangian modified})  when used in Eq. (\ref{Mu_matrix_equation}) yield the light neutrino mass matrix as:
\begin{equation}
  m_{\nu}=F_{2}\begin{bmatrix}
2\upsilon^{\dagger}_{s}\Phi_{a} + \upsilon^{\dagger}_{\kappa} & -\upsilon^{\dagger}_{s}\Phi_{c} + \upsilon^{\dagger}_{\varphi} & -\upsilon^{\dagger}_{s}\Phi_{b} + \upsilon^{\dagger}_{\zeta} \\
-\upsilon^{\dagger}_{s}\Phi_{c}+ \upsilon^{\dagger}_{\varphi}  & 2\upsilon^{\dagger}_{s}\Phi_{b}+ \upsilon^{\dagger}_{\zeta} & -\upsilon^{\dagger}_{s}\Phi_{a}+ \upsilon^{\dagger}_{\kappa}   \\
-\upsilon^{\dagger}_{s}\Phi_{b}+ \upsilon^{\dagger}_{\zeta} & -\upsilon^{\dagger}_{s}\Phi_{a}+ \upsilon^{\dagger}_{\kappa} & 2\upsilon^{\dagger}_{s}\Phi_{c} +  \upsilon^{\dagger}_{\varphi} 
\end{bmatrix}\textrm{,}
\label{LSS:final matrix with vev}
\end{equation}
\vspace{0.1 in}
where, $ F_{2}=\dfrac{2 Y_{L}Y_{D}\upsilon^{2}_{h}\upsilon^{\dagger}_{\varepsilon}}{\Lambda^{2}Y_{M}\upsilon_{\varphi^{\prime}}} $ is a dimensionless constant.

\section{Numerical analysis}
\label{sec:num}
The global-fit analysis constrains the non-unitary parameter \cite{Fernandez-Martinez:2016lgt, Wang:2021rsi} as\\
\begin{equation}
\vert \mathcal{\eta} \vert < \left(\begin{matrix}
 1.25 \times 10^{-3} & 1.20 \times 10^{-5}  & 1.35 \times 10^{-3} \\
 1.20 \times 10^{-5} & 2.21 \times 10^{-4} & 6.13 \times 10^{-4} \\
1.35 \times 10^{-3}  & 6.13 \times 10^{-4} & 2.81 \times 10^{-3}
\end{matrix}\right)\textrm{.}
\label{eta matrix}
\end{equation}
Since both of our $ A_{4} $ symmetry based inverse and linear seesaw  models predict $ M_{D}M^{-1} $ to be a diagonal matrix, from Eqn. (\ref{eta matrix}) we choose the strongest experimental bound of $ \eta $ as the constraint for our models,  i.e.,  $\eta  < 2.21 \times 10^{-4}$. We randomly choose four different values of $ \eta$ which satisfy the constraint $ < {\mathcal{O}}(10^{-4})$ and use it to find the branching ratio of $ BR(\mu \rightarrow e + \gamma )$ for the three allowed VEV alignments of the triplet flavon $\Phi_s$ (0,1,1) (NH), (-1,1,1) (NH) and (0,1,-1) (IH). It may be noted that in our previous work \cite{Devi:2021aaz}, only for these three cases we had obtained the neutrino oscillation parameters within the 3$\sigma$ ranges of their current allowed global best fit values. We do the analysis for four randomly chosen (and allowed) values of the non-unitarity parameter  $ \eta = 2.19 \times 10^{-4}$, $ \eta = 4.0 \times 10^{-6}$, $ \eta = 5.0 \times 10^{-7}$ and $ \eta = 9.0 \times 10^{-9}$.  \\
 \begin{figure*}
  \centering
\includegraphics[width=0.7\textwidth]{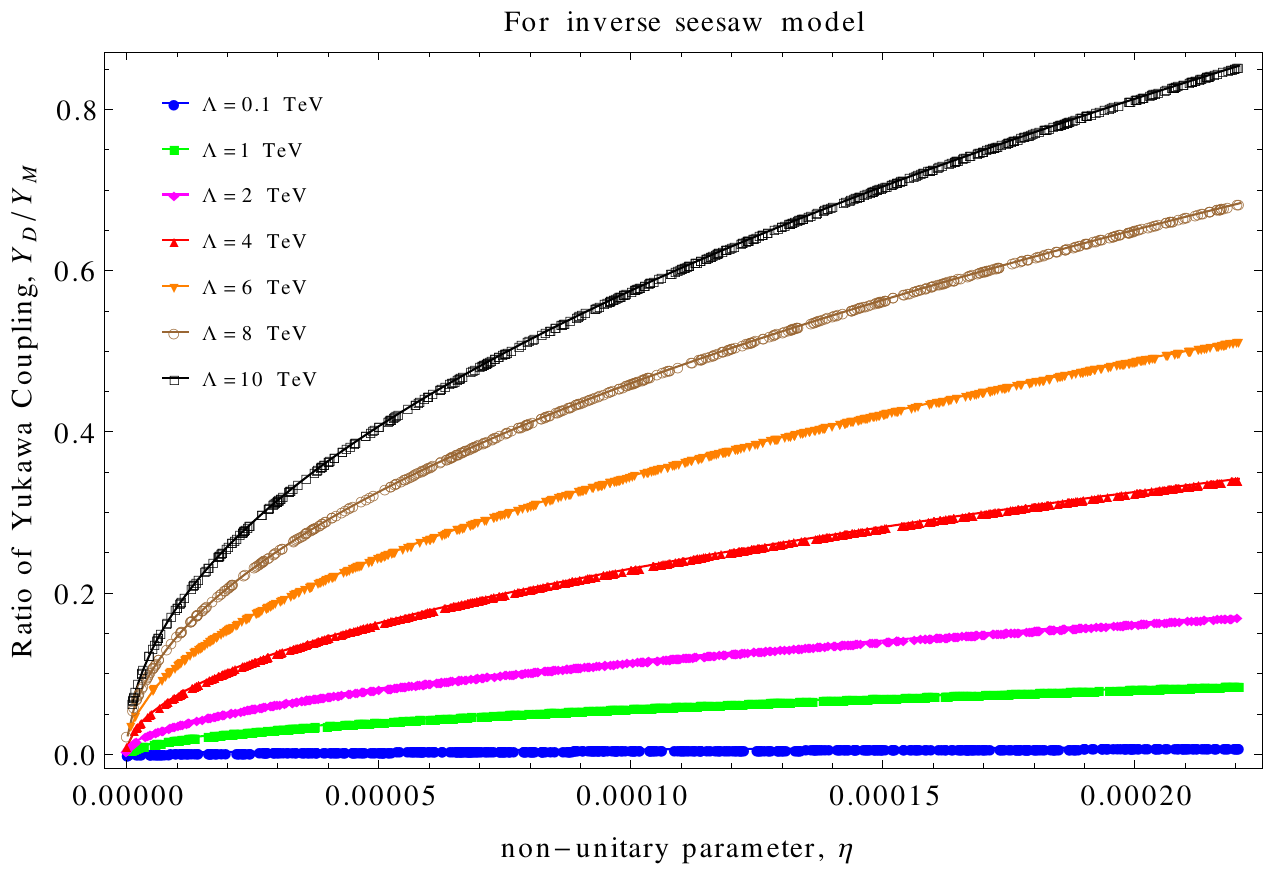} 
 \caption{\textit{ Correlation plots between $ a $ and non-unitary parameter $ \eta $ for different values of $ \Lambda $ }}
 \label{Lambda}
\end{figure*}


\begin{table}
 \begin{center}
 \scalebox{0.8}{
\begin{tabular}{| *{8}{c|} }
    \hline
SL. NO.  &  VEV   & \multicolumn{2}{c|}{ISS/LSS}
            & SL. NO.  &  VEV   & \multicolumn{2}{c|}{ISS/LSS}
              \\
    \hline
     &    &   NH  &   IH   &       &    &   NH  &   IH   \\
    \hline
1   &  (1,0,0)  & $ \textbf{\textcolor{green}{\sffamily X}} $  &    $ \textbf{\textcolor{green}{\sffamily X}} $  & 14   &  (-1,1,-1)  &  $ \textbf{\textcolor{green}{\sffamily X}} $     &  $ \textbf{\textcolor{green}{\sffamily X}} $       \\
    \hline
2   &  (0,1,0)  & $ \textbf{\textcolor{green}{\sffamily X}} $  &    $ \textbf{\textcolor{green}{\sffamily X}} $   &  15   &  (-1,-1,1)  &   $ \textbf{\textcolor{green}{\sffamily X}} $  &    $ \textbf{\textcolor{green}{\sffamily X}} $   \\
    \hline
3   &  (0,0,1)  &  $ \textbf{\textcolor{green}{\sffamily X}} $  &    $ \textbf{\textcolor{green}{\sffamily X}} $  &   16   &  (1,1,-1)  &  $ \textbf{\textcolor{green}{\sffamily X}} $  &    $ \textbf{\textcolor{green}{\sffamily X}} $   \\
    \hline
4   &  (1,1,0)  &   $ \textbf{\textcolor{green}{\sffamily X}} $  &    $ \textbf{\textcolor{green}{\sffamily X}} $   &  17   &  (1,-1,1)  & $ \textbf{\textcolor{green}{\sffamily X}} $  &    $ \textbf{\textcolor{green}{\sffamily X}} $   \\
    \hline
5   &  (1,0,1)  & $ \textbf{\textcolor{green}{\sffamily X}} $  &    $ \textbf{\textcolor{green}{\sffamily X}} $   &  18   &  (-1,1,1)  &       allowed &   $ \textbf{\textcolor{green}{\sffamily X}} $   \\
    \hline
6   &  (0,1,1)  & allowed      &  $ \textbf{\textcolor{green}{\sffamily X}} $    &    19   &  (-1,-1,0)  &  $ \textbf{\textcolor{green}{\sffamily X}} $  &    $ \textbf{\textcolor{green}{\sffamily X}} $   \\
    \hline
7   &  (1,-1,0)  &  $ \textbf{\textcolor{green}{\sffamily X}} $  &    $ \textbf{\textcolor{green}{\sffamily X}} $    &  20   &  (-1,0,-1)  &    $ \textbf{\textcolor{green}{\sffamily X}} $  &    $ \textbf{\textcolor{green}{\sffamily X}} $   \\
    \hline
8   &  (1,0,-1)  &   $ \textbf{\textcolor{green}{\sffamily X}} $  &    $ \textbf{\textcolor{green}{\sffamily X}} $    &   21   &  (0,-1,-1)  &allowed  &    $ \textbf{\textcolor{green}{\sffamily X}} $    \\
    \hline
9   &  (-1,0,1)  &  $ \textbf{\textcolor{green}{\sffamily X}} $  &    $ \textbf{\textcolor{green}{\sffamily X}} $  &   22   &  (-1,0,0)  &   $ \textbf{\textcolor{green}{\sffamily X}} $  &    $ \textbf{\textcolor{green}{\sffamily X}} $       \\
    \hline
10   &  (-1,1,0)  & $ \textbf{\textcolor{green}{\sffamily X}} $  &    $ \textbf{\textcolor{green}{\sffamily X}} $  & 23   &  (0,-1,0)  &   $ \textbf{\textcolor{green}{\sffamily X}} $  &    $ \textbf{\textcolor{green}{\sffamily X}} $    \\
    \hline
11   &  (0,-1,1)  &  $ \textbf{\textcolor{green}{\sffamily X}} $   & allowed &    24   &  (0,0,-1)  &  $ \textbf{\textcolor{green}{\sffamily X}} $  &    $ \textbf{\textcolor{green}{\sffamily X}} $   \\
    \hline
12   &  (0,1,-1)  & $ \textbf{\textcolor{green}{\sffamily X}} $   &  allowed    & 25   &  (1,1,1)  & $ \textbf{\textcolor{green}{\sffamily X}} $  &    $ \textbf{\textcolor{green}{\sffamily X}} $    \\
    \hline
13   &  (1,-1,-1)  & allowed   &   $ \textbf{\textcolor{green}{\sffamily X}} $      &   26   &  (-1,-1,-1)  &  $ \textbf{\textcolor{green}{\sffamily X}} $     &   $ \textbf{\textcolor{green}{\sffamily X}} $   \\
    \hline
\end{tabular}}
    \end{center}
    
\caption{ Summary of our result after solving the triplet flavon equations of different VEV alignments for all the inverse and linear seesaw mechanisms. The green sign, $ \textbf{\textcolor{green}{\sffamily X}} $  indicate ruled out cases, where no output is obtained for the tolerance level of $10^{-5} $. }
\label{vevalignmentcases}
\end{table}
\subsection{Numerical analysis for Inverse seesaw model}
\label{num:ISS}

We know that for Inverse seesaw  model \cite{Dev:2009aw, Karmakar:2016cvb}, \\
\begin{equation}
m_{\nu} = M_{D}(M^{T})^{-1}\mu_{s} M^{-1}M^{T}_{D} \Rightarrow m_{\nu} = F\mu_{s} F^{T}
\label{ISSmnu}
\end{equation}
\begin{equation}
\textrm{where, } F = M_{D}M^{-1} \Rightarrow F = \mathbb{C}|I_{3 \times 3}| \textrm{ ; } \mathbb{C}=\dfrac{Y_{D}v_{h}}{\Lambda Y_{M}}
\end{equation}
From Eqn.(\ref{ISSmnu}) we have
\begin{equation}
m_{\nu} = |\mathbb{C}|^{2}\mu_{s} \Rightarrow \mu_{s} = \dfrac{m_{\nu}}{|\mathbb{C}|^{2}}= \dfrac{U_{pmns}m_{\nu_{diag}}U_{pmns}^{T}}{|\mathbb{C}|^{2}}
\label{C1:ISS}
\end{equation}
\begin{equation}
\textrm{Since, } \eta = \dfrac{1}{2}FF^{\dagger} \Rightarrow \eta= \dfrac{1}{2}|\mathbb{C}|^{2}I_{3 \times 3}
\label{eta:ISS1}
\end{equation}
From Eqns.(\ref{C1:ISS}) and (\ref{eta:ISS1}), we can write
\begin{equation}
\mu_{s} =\dfrac{U_{pmns}m_{\nu_{diag}}U_{pmns}^{T}}{2\eta }\textrm{.}
\label{eta:ISS2} 
\end{equation}
Also, for the three allowed cases of VEV alignment of triplet flavon field  $\Phi_s$, the matrix $ \mu_{s} $ from Eqn. (\ref{ISS:Mu}) takes the following different forms:\\
\\
1. For VEV (0,1,1) with normal hierarchy:\\
\begin{equation}
 \mu_{s}=K_{1}\begin{bmatrix}
A_{1} & B_{1}-1 & C_{1}-1\\
 B_{1}-1  & C_{1}+2 & A_{1}\\
C_{1}-1 & A_{1} & B_{1}+2
\end{bmatrix} \textrm{,}
\label{mu:ISS 1}
\end{equation}\\

where, $ K_{1}= \Big\vert \dfrac{Y_{\mu_{s}} v_{\rho^{\prime}}v^{\dagger}_{\rho^{\prime \prime}}v_{s}}{\Lambda^{2}}  \Big\vert$,  $ A_{1} = \Big\vert \dfrac{v_{\Omega}}{v_{s}}  \Big\vert $, $ B_{1}=  \Big\vert \dfrac{v_{\xi}}{v_{s}}  \Big\vert $ and $ C_{1} = \Big\vert \dfrac{v_{\tau}}{v_{s}}  \Big\vert $. \\
\\
2. For VEV (-1,1,1) with normal hierarchy:\\
\begin{equation}
 \mu_{s}=K_{1}\begin{bmatrix}
A_{1}-2 & B_{1}-1 & C_{1}-1\\
 B_{1}-1  & C_{1}+2 & A_{1}+1\\
C_{1}-1 & A_{1}+1 & B_{1}+2
\end{bmatrix} \textrm{,}
\label{mu:ISS 2}
\end{equation} \\
\\
3. For VEV (0,1,-1) with inverted hierarchy:\\
\begin{equation}
 \mu_{s}=K_{1}\begin{bmatrix}
A_{1} & B_{1}+1 & C_{1}-1\\
B_{1}+1  & C_{1}+2 & A_{1}\\
C_{1}-1 & A_{1} & B_{1}-2
\end{bmatrix} \textrm{,}
\label{mu:ISS 3}
\end{equation} \\

where, $ A_{1} $, $ B_{1}$ and $ C_{1} $ take same form in all the three cases above. We use the randomly chosen value of $\eta$ in Eqn. (\ref{eta:ISS2}), and then comparing it with Eqns. (\ref{mu:ISS 1}), (\ref{mu:ISS 2}) and (\ref{mu:ISS 3}), the elements $ A_{1} $, $ B_{1} $ and $C_{1} $ can be computed, with the assumption for simplicity that $ \vert K_{1} \vert  \sim$ 1 eV in further analysis. We take all the non-zero elements of the matrix M in Eqn. (\ref{ISS: MD}) of the order of $ \sim $ 1 TeV. The heavy Majorana  neutrinos $ M_{R} $ which is an admixture of basis N and S have mass eigenvalues given by $ M_{R}=M\pm \dfrac{\mu_{s}}{2} $  \cite{Baldes:2013eva, Dolan:2018qpy} whose values are used in Eqn.(\ref{BR form}). Further, the Dirac Matrix $ M_{D} $ can be constructed, for inverse seesaw mechanism\cite{Forero:2011pc} as,
\begin{equation}
M_{D} =U m^{1/2}_{\nu diag}\mathcal{R}\mu_{s}^{-1/2}M^{T}
\end{equation}
$\mathcal{R}$ being a complex orthogonal matrix satisfying $\mathcal{RR}^{T}=\mathbb{1}_{3\times 3}$. We next calculate B as given in Eqns. (\ref{eqn:K}) and (\ref{B:ISS}) and use the values of $\mathcal{K}  $ from Eqn. (\ref{eqn:K}) to compute the branching ratio in Eqn.(\ref{BR form}). Since $ M_{D}M^{-1}=\left( \dfrac{Y_{D}v}{\Lambda Y_{M}} \right) \mathbb{1}_{3 \times 3} $, from Eqs. (\ref{B:ISS}) and (\ref{non unitarity}), we can write 
\begin{equation}
| \eta |=\dfrac{1}{2}\left[ \dfrac{Y_{D}v_{h}}{\Lambda Y_{M}} \right]^{2}\mathbb{1}_{3 \times 3}\approx \dfrac{a^{2}}{2}\left[ \dfrac{v_{h}}{\Lambda} \right]^{2}\mathbb{1}_{3 \times 3}  
\end{equation}
where $a = \dfrac{Y_{D}}{Y_{M}} $, and  $Y_{D}$ and $Y_{M}$ are Yukawa couplings of the  L-N and N-S sectors respectively. At present, no information is available on $Y_{D}$, $Y_{M}$ or $\Lambda $ and the cut-off scale of (and hence flavour symmetry breaking scale)  the theory  $\Lambda $ can take different possible values. However, constraints on the value of non-unitarity parameter $\eta$ is available, and hence relation in Eqn. (31) can be used to obtain a correlation plot between the parameter \enquote{a} and the non-unitary parameter $ \eta $, which is  shown in Fig (\ref{Lambda}) for different values of the cutoff scale $\Lambda$. This graph depicts that same value  non-unitarity parameter $\eta$ can be obtained for different values  of ratio of couplings $\dfrac{Y_{D}}{Y_{M}} $ if cut-off scale of the theory can be fine tuned. This also implies that same amount of non-unitarity can be generated for different values of scale of flavour symmetry breaking, if $\dfrac{Y_{D}}{Y_{M}} $ can be adjusted accordingly. In other words, it can be stated that all the physical quantities $\Lambda $,$ Y_M, $ and $Y_D$ can not be allowed to change freely in order to generate a given amount of non-unitarity.
 Also, the curve with the steeper slope indicates variation of $ Y_{D}/Y_{M} $ with respect to non-unitary parameter $ \eta $ for a given cut-off scale $ \Lambda $ having higher value than that for the lower curve.  The higher the value of $ \Lambda $, higher is the value of the ratio $ Y_{D}/Y_{M} $  for a given non-unitary parameter within its allowed region of ($ 0< \eta < 2.2099\times 10^{-4} $).  
\subsection{Numerical analysis for Linear seesaw model}
\label{num:LSS}
From  Eqn.(\ref{Mu_matrix_equation}),  it is seen that the mass matrix $ M_{L} $ can be obtained as:
\begin{equation}
\vert M_{L} \vert= \Big\vert \dfrac{m_{\nu}}{ 2M_{D}M^{-1} } \Big\vert
\label{ML:LSS}
\end{equation}
\begin{equation}
\Rightarrow  \vert M_{L} \vert= \Big\vert \dfrac{U_{pmns}m_{\nu diag}U^{T}_{pmns}}{2\sqrt{2 \eta}} \Big\vert
 \textrm{ where, } \eta = \dfrac{1}{2} \Big\vert M_{D}M^{-1}\Big \vert^{2}
 \label{ML:LSS2}
\end{equation}\\
For this model, the $ M_{L} $ mass matrix from Eqn. (\ref{linearmassmatrix2}) for the three allowed vacuum alignments of the triplet flavon $\Phi_s$ can be reduced to the following form:\\

1. For VEV (0,1,1) with normal hierarchy:\\
\begin{equation}
 M_{L}=K_{2}\begin{bmatrix}
A_{2} & B_{2}-1 & C_{2}-1\\
 B_{2}-1  & C_{2}+2 & A_{2}\\
C_{2}-1 & A_{2} & B_{2}+2
\end{bmatrix} \textrm{,}
\label{ML:LSS 1}
\end{equation}
where, $ K_{2}= \Big\vert \dfrac{Y_{L}v_{h}v^{\dagger}_{s}}{\Lambda}  \Big\vert$, $ A_{2} = \Big\vert \dfrac{v^{\dagger}_{\kappa}}{v^{\dagger}_{s}}  \Big\vert $, $ B_{2}=  \Big\vert \dfrac{v^{\dagger}_{\zeta}}{v^{\dagger}_{s}}  \Big\vert $ and $ C_{2} =  \Big\vert \dfrac{v^{\dagger}_{\varphi}}{v^{\dagger}_{s}}  \Big\vert $. \\

2. For VEV (-1,1,1) with normal hierarchy:\\
\begin{equation}
 M_{L}=K_{2}\begin{bmatrix}
A_{2}-2 & B_{2}-1 & C_{2}-1\\
 B_{2}-1  & C_{2}+2 & A_{2}+1\\
C_{2}-1 & A_{2}+1 & B_{2}+2
\end{bmatrix} \textrm{,}
\label{ML:LSS 2}
\end{equation}\\

3.  For VEV (0,1,-1) with inverted hierarchy:\\
\begin{equation}
 M_{L}=K_{2}\begin{bmatrix}
A_{2} & B_{2}-1 & C_{2}+1\\
 B_{2}-1  & C_{2}-2 & A_{2}\\
C_{2}+1 & A_{2} & B_{2}+2
\end{bmatrix} \textrm{,}
\label{ML:LSS 3}
\end{equation}\\

 In Eqn (33), we feed the chosen value fo $\eta$, and then Eqns. (\ref{ML:LSS 1}), (\ref{ML:LSS 2}), (\ref{ML:LSS 3}) are compared with Eqn. (\ref{ML:LSS}) to find the elements $ A_{2} $, $B_{2} $ and $ C_{2} $ by taking $ \vert K_{2}  \vert \sim 1 $ eV for simplicity. We assume all the non-zero elements of M to be $\sim$1 TeV where the heavy Majorana neutrino mass matrix $ M_{R}= \pm M $ \cite{ Dolan:2018qpy}. The Dirac mass matrix $ M_{D} $ is computed  as \cite{Forero:2011pc},\\
\begin{equation}
M_{D}=\vert U m^{1/2}_{\nu diag} \mathcal{R}^{\prime} m^{1/2}_{\nu diag}U^{T}M^{-1}_{L}M^{T} \vert
\end{equation}\\
where, $ \mathcal{R}^{\prime} $ satisfies $ \mathcal{R}^{\prime} + \mathcal{R}^{\prime T}=\mathbb{1}_{3\times 3} $. For further details, one can refer to \cite{Forero:2011pc, Dolan:2018qpy, DelleRose:2015bms}. This is then used in Eqn. (\ref{K3x3}) and  Eqn. (\ref{BR form}) to  compute the branching ratio of $ \mu \rightarrow e + \gamma $ for the four values of non-unitary parameter $ \eta$. We would like to note that correlation plots similar to Fig. 1 can also be obtained for LSS model. Our results based on this analysis are presented and discussed in Section 5.\\

\begin{table}
\centering
{\begin{adjustbox}{width=12cm}
\begin{tabular}{|l|l|l|l|l|l|}\hline
Non-unitary & VEV & NH/IH & Type & Range of $ BR (\mu \rightarrow e + \gamma ) $  & Experiments that can probe \\  
parameter, $ \eta $   &   &   &   & &  \\  \cline{1-6} 
  \multirow{6}{*}{$ 2.19 \times 10^{-4}$ } & \multirow{2}{*}{(0,1,1)} & \textcolor{ForestGreen}{NH}  & \textcolor{OrangeRed}{ISS}   & $ 1.2362 \times 10^{-7} \rightarrow 2.87721 \times 10^{-7} $  &   T, S, E, CB, $ M^{*} $, M, MII, NG\\  \cline{3-6}  
  & & \textcolor{ForestGreen}{NH}  &  \textcolor{TealBlue}{LSS}    &  $ 1.00088 \times 10^{-10} \rightarrow 6.51129\times 10^{-10} $  &   E, CB, $ M^{*} $, M, MII, NG 
    \\   \cline{2-6}
   & \multirow{2}{*}{(-1,1,1)} & \textcolor{ForestGreen}{NH} & \textcolor{OrangeRed}{ISS} & $  9.82583\times 10^{-9} \rightarrow 5.50333\times 10^{-7} $    & T, S, E, CB, $ M^{*} $, M, MII, NG    \\ \cline{3-6}
  & & \textcolor{ForestGreen}{NH}  &   \textcolor{TealBlue}{LSS} &  $  9.05723\times 10^{-13} \rightarrow 1.39745\times 10^{-12} $ &  M, MII, NG 
   \\ \cline{2-6}
 & \multirow{2}{*}{(0,1,-1)} & \textcolor{BurntOrange}{IH} & \textcolor{OrangeRed}{ISS} &   $ 2.8648 \times 10^{-9} \rightarrow 9.51365 \times 10^{-8} $ &  T, S, E, CB, $ M^{*} $, M, MII, NG  \\ \cline{3-6}
  & & \textcolor{BurntOrange}{IH}  &   \textcolor{TealBlue}{LSS} &  $ 6.54537 \times 10^{-8} \rightarrow 1.19193 \times 10^{-7} $ &  T, S, E, CB, $ M^{*} $, M, MII, NG 
   \\ \cline{2-6}
    \cline{1-6}
   \multirow{6}{*}{$4.0 \times 10^{-6}$ } & \multirow{2}{*}{(0,1,1)} & \textcolor{ForestGreen}{NH} & \textcolor{OrangeRed}{ISS} &    $  4.39094 \times 10^{-11} \rightarrow 9.32741\times 10^{-11} $  & CB, $ M^{*} $, M, MII, NG    \\ 
\cline{3-6}
  & & \textcolor{ForestGreen}{NH} &   \textcolor{TealBlue}{LSS} &  $  1.2099 \times 10^{-12} \rightarrow  5.33457\times 10^{-12} $  & M, MII, NG
   \\ \cline{2-6}
   & \multirow{2}{*}{(-1,1,1)} & \textcolor{ForestGreen}{NH} & \textcolor{OrangeRed}{ISS} &  $ 2.54968 \times 10^{-12} \rightarrow 1.47428\times 10^{-10} $   &   CB, $ M^{*} $, M, MII, NG   \\ \cline{3-6}
  & & \textcolor{ForestGreen}{NH} &   \textcolor{TealBlue}{LSS} &  $ 1.12946 \times 10^{-12} \rightarrow 2.1398\times 10^{-12} $ & $ M^{*} $, M, MII, NG 
   \\ \cline{2-6}
 & \multirow{2}{*}{(0,1,-1)}  & \textcolor{BurntOrange}{IH} & \textcolor{OrangeRed}{ISS} &   $  7.78198\times 10^{-13} \rightarrow 3.21865\times 10^{-11} $  &    $ M^{*} $, M, MII, NG   \\ \cline{3-6}
  & & \textcolor{BurntOrange}{IH}  &   \textcolor{TealBlue}{LSS} &  $ 3.41281 \times 10^{-11} \rightarrow 4.84894\times 10^{-11} $ &  $ M^{*} $, M, MII, NG 
   \\ \cline{2-6}
    \cline{1-6}
     \multirow{6}{*}{$  5.0 \times 10^{-7}$ } & \multirow{2}{*}{(0,1,1)} & \textcolor{ForestGreen}{NH} & \textcolor{OrangeRed}{ISS} &    $  6.87031 \times 10^{-13} \rightarrow 1.45709 \times 10^{-12} $  &  M, MII, NG\\ \cline{3-6}
  & & \textcolor{ForestGreen}{NH}  &   \textcolor{TealBlue}{LSS} &  $ 9.8222 \times 10^{-13} \rightarrow 3.17343 \times 10^{-11} $ & $ M^{*} $, M, MII, NG 
   \\ \cline{2-6}
   & \multirow{2}{*}{(-1,1,1)} & \textcolor{ForestGreen}{NH} & \textcolor{OrangeRed}{ISS} &  \textcolor{Thistle}{$ 4.04261 \times 10^{-14} \rightarrow 2.29969\times 10^{-12} $ }    &   M, MII, NG   \\ \cline{3-6}
  & & \textcolor{ForestGreen}{NH} &   \textcolor{TealBlue}{LSS} &  $ 7.6931 \times 10^{-12} \rightarrow 7.90733 \times 10^{-11} $ & CB, $ M^{*} $, M, MII, NG 
   \\ \cline{2-6}
 & \multirow{2}{*}{(0,1,-1) }  & \textcolor{BurntOrange}{IH} & \textcolor{OrangeRed}{ISS} &  \textcolor{Thistle}{ $ 1.2138 \times 10^{-14} \rightarrow 5.03077\times 10^{-13} $}  &  M, MII, NG    \\ \cline{3-6 }
  & & \textcolor{BurntOrange}{IH} &   \textcolor{TealBlue}{LSS} &  $ 9.23256 \times 10^{-13} \rightarrow 1.42411 \times 10^{-12} $ &   M, MII, NG 
   \\ \cline{2-6}
    \cline{1-6}
     \multirow{6}{*}{$ 9.0 \times 10^{-9}$ } & \multirow{2}{*}{(0,1,1)} & \textcolor{ForestGreen}{NH} & \textcolor{OrangeRed}{ISS} &    $  2.22641\times 10^{-16} \rightarrow 4.72084 \times 10^{-16} $  &  NG\\ \cline{3-6}
  & & \textcolor{ForestGreen}{NH} &   \textcolor{TealBlue}{LSS} &  $ 7.77436 \times 10^{-15} \rightarrow 1.94312 \times 10^{-17} $ & NG
   \\ \cline{2-6}
   & \multirow{2}{*}{(-1,1,1)} & \textcolor{ForestGreen}{NH} & \textcolor{OrangeRed}{ISS} &   $  1.3125\times 10^{-17} \rightarrow 7.44927\times 10^{-17} $  &  NG \\ \cline{3-6}
  & & \textcolor{ForestGreen}{NH} &   \textcolor{TealBlue}{LSS} &  $ 3.357874 \times 10^{-15} \rightarrow 8.46285 \times 10^{-15} $ & NG
   \\ \cline{2-6}
 & \multirow{2}{*}{(0,1,-1)}  & \textcolor{BurntOrange}{IH} & \textcolor{OrangeRed}{ISS} &  $ 3.93177 \times 10^{-18} \rightarrow 1.63004 \times 10^{-16} $  & NG  \\ \cline{3-6}
  & & \textcolor{BurntOrange}{IH} &   \textcolor{TealBlue}{LSS} &  $ 3.32112 \times 10^{-16} \rightarrow 5.74934 \times 10^{-16} $ & NG
   \\ \cline{2-6}
    \cline{1-6}
\end{tabular}
\end{adjustbox}}
\caption{\textit{Results of this work on Branching Ratio of the cLFV decay $ \mu \rightarrow e + \gamma $ for allowed vacuum alignments of the triplet flavon, for ISS and LSS models and the different experiments that can probe them. Here NH and IH represents normal and inverted hierarchies respectively. The color codes indicates  \textcolor{ForestGreen}{$ \blacksquare $} $ \rightarrow $ NH (Normal Hierarchy), \textcolor{BurntOrange}{$ \blacksquare $} $ \rightarrow $ IH (Inverted Hierarchy), 
 \textcolor{OrangeRed}{$ \blacksquare $} $ \rightarrow $ ISS (Inverse Seesaw), \textcolor{TealBlue}{$ \blacksquare $}  $ \rightarrow $ LSS (Linear Seesaw). \textcolor{Thistle}{$ \blacksquare $} represents the dataset for which the upper limit of  $BR(\mu \rightarrow e + \gamma) $ falls in the current limit of MEG and future sensitivity of MEG II. Here the experiments are indexed as T for TRIMUF, S for SIN, E for E328, CB for Crystal Box, $ M^{*} $ for MEGA, M for MEG, MII for MEG II and NG for next generation muon beam experiments such as AMF in Fermilab. The experiments such as TRIUMF, SIN, E328, Crystal Box and MEGA are not operating currently and MEG and MEG II are the currently active experiments. Here NG means next-generation experiments.}}
\label{Tab:3}
\end{table}

\subsection{Dynamics of flavour symmetry}
\label{K1,K2}
For ISS model as given in section (\ref{subsec:ISS}), we can consider the typical energy scales of the different mass matrices as - $ M_{D} =10$ GeV, $ M =1$ TeV, $ v_{h} =246$ GeV. This gives $\dfrac{v_{h}}{\Lambda}= 2.46 \times 10^{-4} \approx 2.5 \times 10^{-4} $  for a cut-off scale of $ \Lambda =10^{3} $ TeV. From Eqns.(\ref{ISS: MD}) and (\ref{ISS:Mu}), we can write:
\begin{equation}
\dfrac{M_{D}}{M}= \left( \dfrac{Y_{D}}{Y_{M}} \right) \left( \dfrac{v_{h}}{\Lambda} \right)\approx 10^{-2}
\end{equation} 
and we get
\begin{equation}
  Y_M = 0.025 Y_D  
\end{equation}
Using the chosen values $ K_{1}=1$ eV  $ Y_{\mu_{s}} \sim 0.01 $, one gets\\
\begin{equation}
v_{\rho^{\prime}}v^{\dagger}_{\rho^{\prime \prime}}v_{s}= 10^{-4}  (\textrm{ TeV } )^{3}
\end{equation}
\\
This condition can be satisfied by taking a set of values for the flavon VEVs and coupling constants as shown in Table (\ref{tab:a}) without affecting the overall gauge symmetries and interactions considered in the model. 
For this set of values we get $A_{1}\sim 2000$, $ B_{1}\sim 1500$ and $ C_{1}\sim 1000 $ which agree with some typical values of $ A_{1} $, $ B_{1} $ and $ C_{1} $ from their actual data obtained in our computation. Proceeding similarly for the LSS model, considering the energy scales $M_{D}=10 $ GeV,  $ M =1$ TeV, $ M_{L} =10$ eV and $ \Lambda =10^{3} $ TeV , we can evaluate from Eqns. (\ref{linearmassmatrix1}) and (\ref{linearmassmatrix2}) as 
\begin{equation}
\dfrac{M_{D}}{M}=10^{-2}\Rightarrow \dfrac{Y_{D}\upsilon^{\dagger}_{\varepsilon}}{Y_{M}\upsilon_{\varphi^{\prime}}}\times 2.5 \times 10^{-2}=1 
\end{equation}
\\
and it can be shown that
\begin{equation}
\dfrac{\upsilon^{\dagger}_{\varepsilon} }{\upsilon_{\varphi^{\prime}} }>1
\label{Ineq3}
\end{equation}
\\
Also, using the chosen value $ K_{2}=1 $ eV in section (\ref{subsec:LSS}), we get
\begin{equation}
 K_{2}=\dfrac{Y_{L}v_{h}v^{\dagger}_{s}}{\Lambda}\Rightarrow Y_{L}v_{h}v^{\dagger}_{s} = 4  \textrm{ keV }   \textrm{, (where } \Lambda = 10^{3} \textrm{ TeV} )
\label{Ineq4}
\end{equation}
\\
 Eqns  (\ref{Ineq3}) and (\ref{Ineq4}) can be satisfied by choosing a set of values as shown in Table (\ref{tab:b}). Using these values, we obtain $ A_{2}=B_{2}=C_{2}=10$ which agrees with of the values of $ A_{2} $, $ B_{2} $ and $ C_{2}$ obtained from computation.\\

 For above set of scales and couplings,  $m_\nu$ is obtained in the sub-eV range. For  LSS, the constants $A_2$, $B_2$ and $C_2$ are computed to be of the order of $10$, two orders of magnitude smaller than those of ISS model. It may be noted that we showed this analysis for scales and couplings for some chosen values for demonstration purpose, and can be done for their other values also, such that they satisfy various constraints. Hence, it is seen that flavons corresponding to different representations of $A_4$ group obtain VEVs across  a range of scales, and the flavour symmetry breaking exhibits a very rich and dynamic structure in the two models. Also,  it is seen that \\
 
  \begin{equation}
 \dfrac{\mu_S(=1 keV)}{M_L(=10 eV)} \sim \dfrac{A_1}{A_2} \sim \dfrac{B_1}{B_2} \sim \dfrac{C_1}{C_2} \sim100
 \end{equation}\\
 which are required to obtain similar value of light neutrino mass 0.1 eV in the two models. It should be remembered that $\mu_s$ and $M_L$ correspond to lepton number breaking scale in ISS and LSS respectively (and hence should be small). Moreover, for the purpose of comparison between the two models, we have chosen same values for  $M_D$, $M$, $Y_M$, $Y_D$, $\Lambda$, and $m_\nu$ in Table 5.

\begin{table}
    \begin{minipage}[h]{0.4\textwidth}
        \centering
        \begin{tabular}{| l | l | }
    \hline 
   $\mu_s$   &  1 keV \\
        \hline 
  $M_D$   & 10 GeV \\
        \hline 
   $ M $  & 1 TeV\\
        \hline 
  $ y_M $   & 0.00125 \\
        \hline 
 $ y_D $     & 0.05\\
        \hline 
  $y_{\mu_s} $   & 0.01\\
        \hline 
  $\Lambda$    & $10^3$ TeV\\
        \hline 
  $<v_s>$   & 0.05 TeV\\
        \hline 
 $<v_{\rho'}>$    & 0.05 TeV \\
        \hline 
 $<v_{\rho''}>$    & 0.05 TeV\\
        \hline 
  $<v_{\Omega}>$   & 100 TeV\\
        \hline 
    $<v_{\zeta}>$  & 75 TeV \\
        \hline 
   $<v_{\tau}>$  & 50 TeV \\
        \hline 
       \end{tabular}
       \caption*{\textit{(a) Scales, Yukawa couplings and various flavon VEVs in a case in ISS obtained in this work, for $K_1$ = 1 eV, $m_\nu = 0.1$ eV. Typical values as obtained in our computation $A_1$=2000, $ B_1$=1500, $C_1$=1000 have been used to estimate various flavon VEVs.}}
       \label{tab:a}
    \end{minipage}
    \hfill
    \begin{minipage}[h]{0.4\textwidth}
        \centering
                \begin{tabular}{| l | l | }
    \hline 
 $M_L$   & 10  eV \\
        \hline 
    $M_D$ & 10 GeV \\
        \hline 
   $ M $  & 1 TeV\\
        \hline 
     $ y_M $  & 0.00125\\
        \hline 
    $ y_D $ & 0.05 \\
        \hline 
  $y_L $   & 0.01\\
        \hline 
   $\Lambda$    & $10^3$ TeV \\
        \hline 
   $<v_s^{\dagger} >$  & 0.4 MeV \\
        \hline 
  $<v_{\epsilon^\dagger}>$   & 800 TeV\\
        \hline 
  $<v_{\kappa^{\dagger}}>$   & 4 MeV\\
        \hline 
  $<v_{\zeta^{\dagger}}>$   &  4 MeV \\
        \hline 
  $<v_{\varphi^{\dagger}}>$   & 4 MeV\\
        \hline 
  $<v_\varphi^{\prime}>$   & 800 TeV \\
        \hline 
        \end{tabular}
        \caption*{\textit{ (b) Scales, Yukawa couplings and various flavon VEVs in a case in LSS obtained in this work, for $K_2$ = 1 eV, $m_\nu = 0.1$ eV. Typical values $A_2= B_2= C_2=10$ as obtained in computation have been used to estimate various flavon VEVs.}}
        \label{tab:b}
     \end{minipage}
     \caption{}
     \label{tab:temps}
\end{table}

\section{Results and Discussion}
\label{sec:result} 
The branching ratio of $ \mu \rightarrow e + \gamma $ for four randomly chosen and allowed values of non-unitary parameter $ \eta $,  for the three allowed vacuum alignments for ISS and LSS models, is computed for seesaw scale $\sim$ 1 TeV. The seesaw scale plays significant role in BR of muon decay, and and hence it is possible to obtain different value of the BR for a different value of seesaw scale. For simplicity, we used $K_1\sim K_2 \sim 1$ eV. The results are shown in Figs. (2-4) and Tables 4 and 5, which can guide that which of the case with a particular flavon VEV and mass hierarchy can be tested or eliminated by current bounds and future sensitivities of  various $BR(\mu \rightarrow e + \gamma)$ experiments. In Figs. (2-4), we have shown the BR results of only those cases that give values of BR allowed by bounds of MEG, and sensitivity limits of MEG II and NG (next generation) experiments. The following observations are in order :\\
\\
For same values of $m_\nu$, seesaw scale, Dirac mass $M_D$, Yukawas, cut-off scale $\Lambda$ (as explained in section 4.3),  from the results in Tables 4-5 and Figs. (2-4), it is seen that:\\

1. The BR($ \mu \rightarrow e + \gamma $) in both the seesaw models depends on the chosen value of the non-unitarity parameter $\eta$, triplet flavonVEV alignment and MH of light neutrinos.\\

2. For higher  values of $\eta$, the BR in LSS is generally smaller/larger than that in ISS for NH/IH case, and while for smaller values of $\eta$, the BR in LSS is larger than that in ISS for both hierarchies.\\

3. It is observed that the BR as computed in our work, for $\eta=5.0\times 10^{-7}$, VEV of $\Phi_s$(-1,1,1)/NH and (0,1,-1)/IH for ISS show closest agreement with the current bounds of MEG and sensitivity limits of MEG II experiment, and this validates our model. Hence, when neutrino MH is fixed in future, one may pinpoint the favoured VEV alignment of triplet flavon $\phi_s$.\\

4. However, the BR for lower values of  $\eta=9.0\times 10^{-9}$ will be testable at next generation experiments only, as their BR values lie beyond the sensitivity limits of planned experiment like MEG II as well. Hence our predictions for these cases may be tested at next generation experiments.\\

5. When light neutrino mass and hierarchy is fixed in future by some other experiments, then through our results presented here, it would be possible to discriminate among the two models, i.e. which one out of LSS/ISS would be more favorable as preferred by results of cLFV experiments. It would also be possible to pinpoint the favorable VEV alignment of the triplet flavon. \\

6. Flavon VEVs, and (hence flavour symmetry breaking scale) show a rich variation of scales, as is seen from results in Table 5, and will be testable in future experiments.\\

\begin{figure}[h]  
\centering
\includegraphics[width=0.6\textwidth]{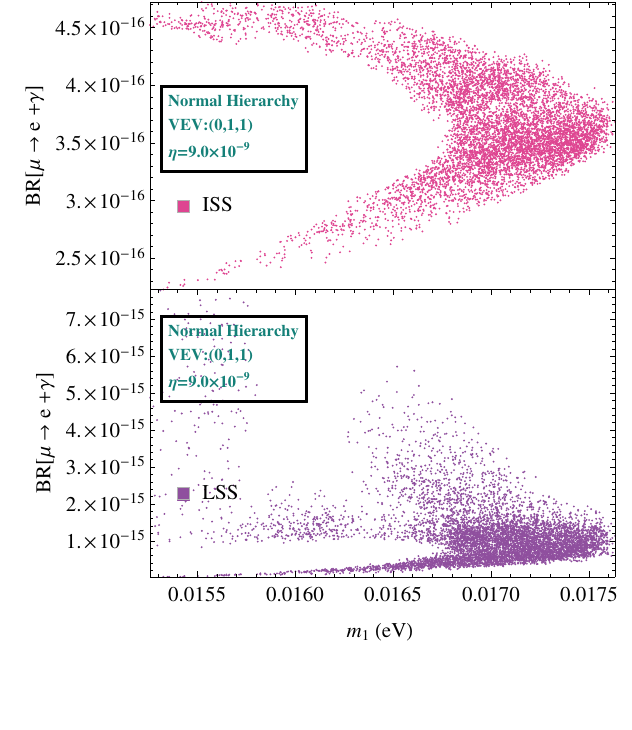}
 \vspace{-1\baselineskip}
 \caption{\textit{Correlation plot between $ BR(\mu \rightarrow e + \gamma ) $ and non-unitary parameter $ \eta $ for allowed vacuum alignment (0,1,1),  $ \Phi_{s} $ of triplet scalar field involved in both the inverse and linear seesaw models with normal hierarchy for non-unitary parameter $ \eta =  9.0 \times 10^{-9}$. These BR values will be testable at NG experiments.}}
\label{Fig011}
\end{figure}

\begin{figure}[h!]  
\centering
    \begin{minipage}[b]{\textwidth}   
     \centering  
   \includegraphics[width=0.46\textwidth]{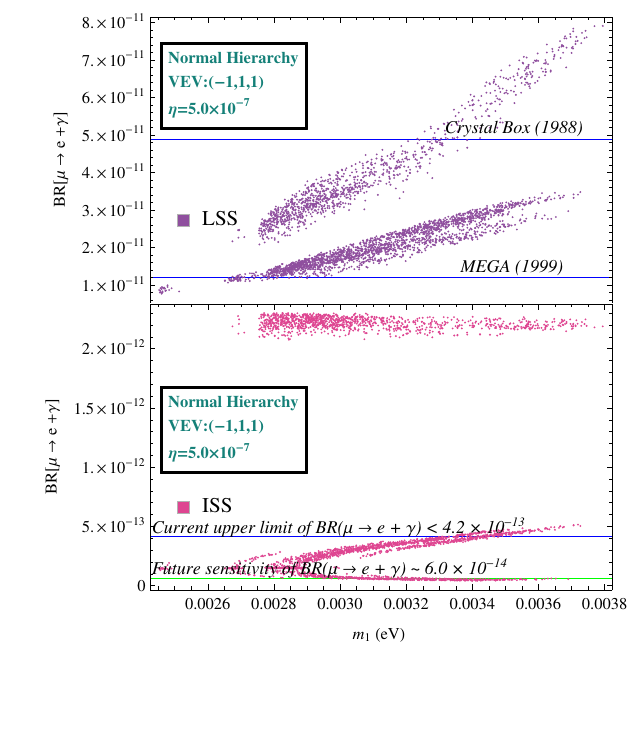} 
   \vspace{0.4\baselineskip}
    \caption*{(a)}
       \hspace{6pt}
         \label{fig:y equals x}
     \end{minipage} 
  
    \begin{minipage}[b]{\textwidth}
         \centering
   \includegraphics[width=0.46\textwidth]{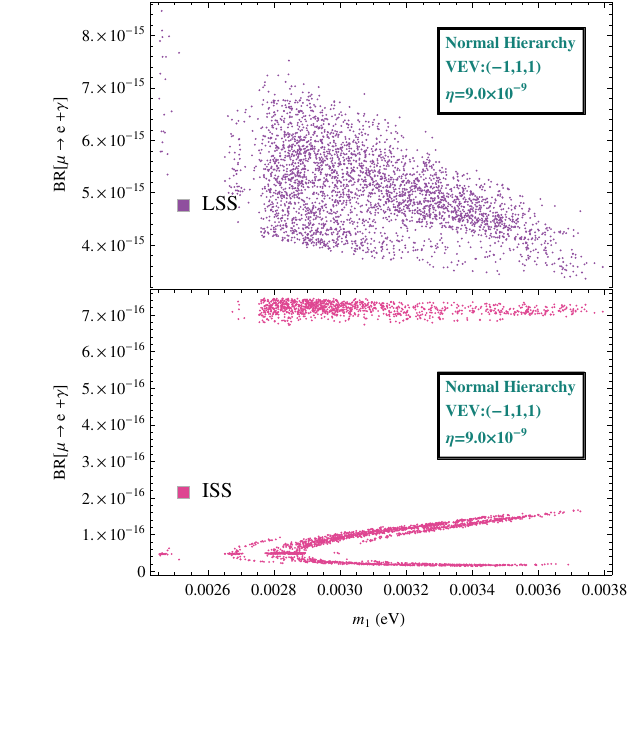}
    \vspace{-2\baselineskip}
    \caption*{(b)}
         \label{fig:three sin x}
     \end{minipage}
      \vspace{-1\baselineskip}
    \caption{\textit{{\footnotesize Correlation plot between $ BR(\mu \rightarrow e + \gamma) $ and non-unitary parameter $ \eta $ for allowed vacuum alignment (-1,1,1) of triplet scalar field, $ \Phi_{s} $  involved in (a) inverse seesaw model for normal hierarchy with  non-unitary parameter, $ \eta= 5.0 \times 10^{-7}$, these BR values will be testable at MEG and MEG II  experiments  and (b) for both the inverse and linear seesaw models with normal hierarchy for $ \eta= 9.0 \times 10^{-9} $, these BR values will be testable at NG experiments.}}}
      \label{Fig-111}
\end{figure}
\begin{figure}[h]  
\centering
    \begin{minipage}[b]{\textwidth}   
     \centering  
   \includegraphics[width=0.46\textwidth]{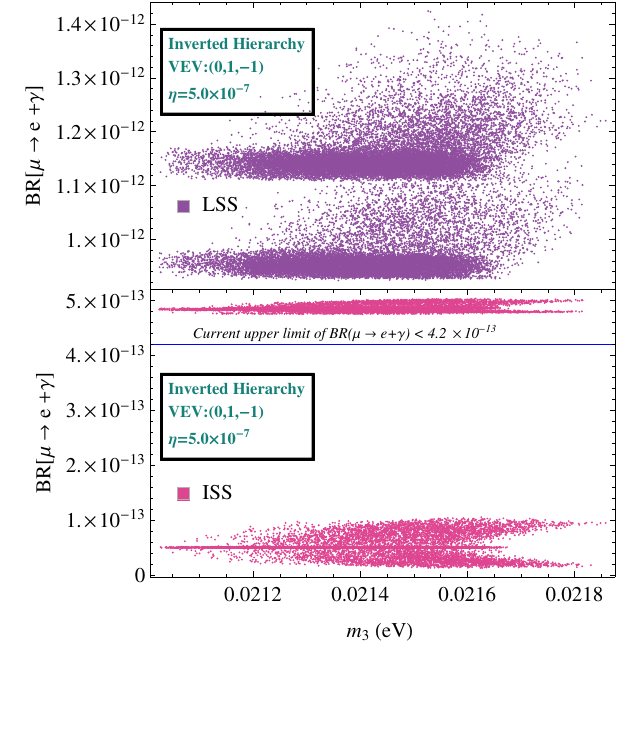} 
   \vspace{0.5\baselineskip}
    \caption*{(a)}
       \hspace{10pt}
         \label{fig:y equals x}
     \end{minipage} 
  
    \begin{minipage}[b]{\textwidth}
         \centering
   \includegraphics[width=0.46\textwidth]{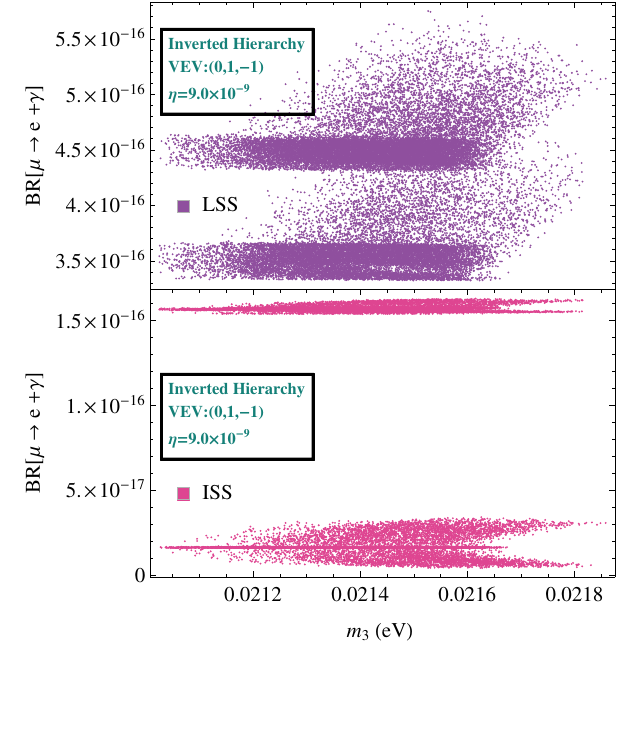}
  \vspace{-2\baselineskip}
    \caption*{(b)}
         \label{fig:three sin x}
     \end{minipage}
    \vspace{-1\baselineskip}
    \caption{\textit{{\footnotesize Correlation plot between $ BR(\mu \rightarrow e + \gamma) $ and non-unitary parameter $ \eta $ for allowed vacuum alignment (0,1,-1) of triplet scalar field, $ \Phi_{s} $  involved in (a) inverse seesaw model for inverted hierarchy with  non-unitary parameter, $ \eta= 5.0 \times 10^{-7}$, these BR values lie within allowed limits of MEG experiment  and (b) for both the inverse and linear seesaw models with inverted hierarchy for $ \eta= 9.0 \times 10^{-9} $, these BR values will be testable at NG experiments.}}}
        \label{Fig01-1}
\end{figure}

\begin{figure}[h]  
\centering
    \begin{subfigure}[b]{\textwidth}   
     \centering  
   \includegraphics[width=0.46\textwidth]{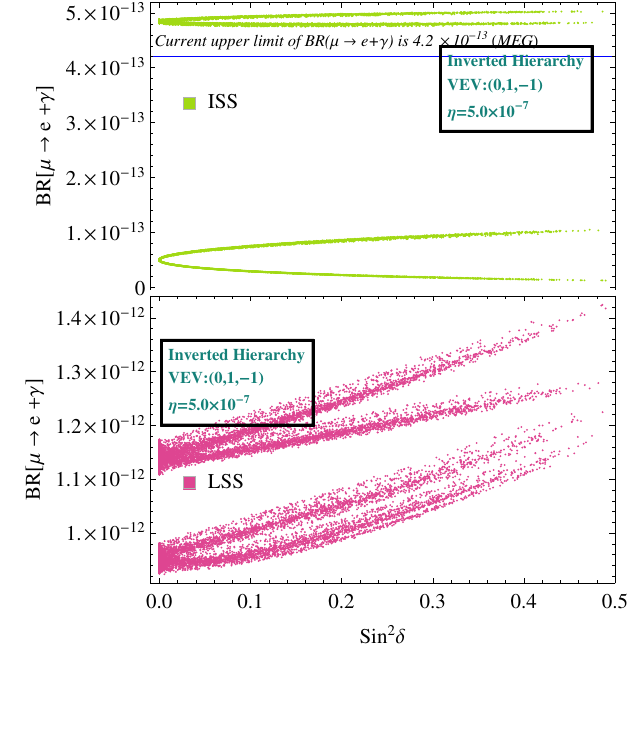} 
   \vspace{-0.001\baselineskip}
    \caption{}
       \hspace{10pt}
         \label{}
     \end{subfigure} 
  
    \begin{subfigure}[b]{\textwidth}
         \centering
   \includegraphics[width=0.46\textwidth]{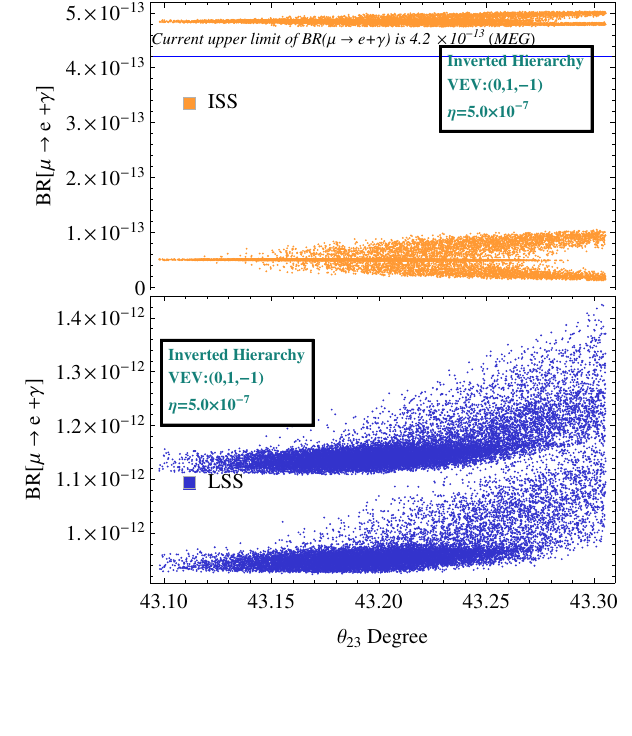}
  \vspace{-0.001\baselineskip}
    \caption{}
         \label{}
     \end{subfigure}
      \hspace{2px}
    \caption{{\footnotesize (a) Correlation plot between $ BR(\mu \rightarrow e + \gamma) $ and $ Sin^{2}\delta $ and (b) Correlation plot between $ BR(\mu \rightarrow e + \gamma) $ and $ \theta_{23} $  for allowed vacuum alignment (0,1,-1) of triplet scalar field, $ \Phi_{s} $  involved in  both the inverse and linear seesaw models with inverted hierarchy with  non-unitary parameter, $ \eta= 5.0 \times 10^{-7}$.}}
        \label{Fig01-1:4c}
\end{figure}

\begin{figure}[h]  
\centering
    \begin{subfigure}[b]{\textwidth}   
     \centering  
   \includegraphics[width=0.46\textwidth]{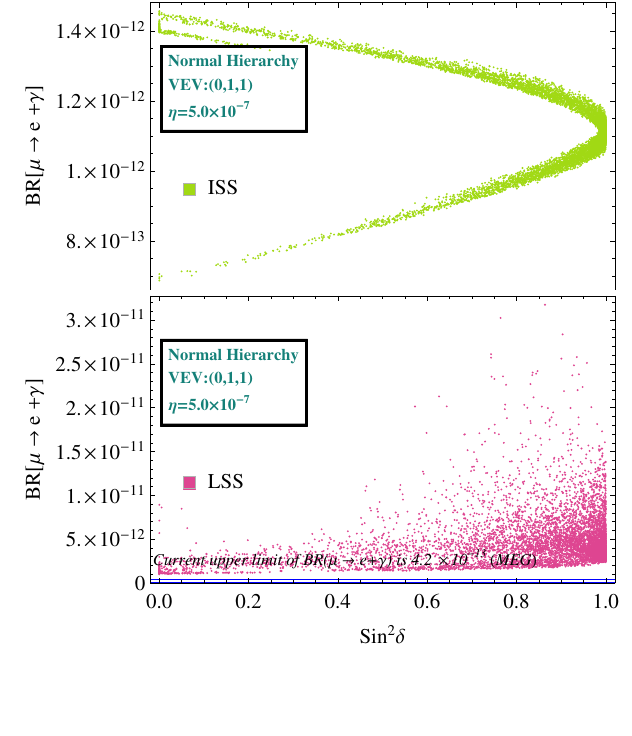} 
   \vspace{-0.001\baselineskip}
    \caption{}
       \hspace{10pt}
         \label{}
     \end{subfigure} 
  
    \begin{subfigure}[b]{\textwidth}
         \centering
   \includegraphics[width=0.46\textwidth]{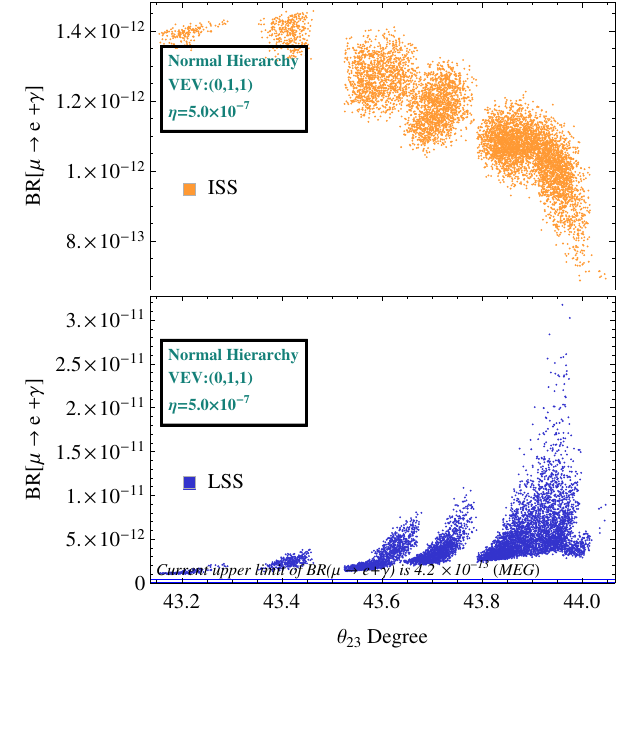}
  \vspace{-0.001\baselineskip}
    \caption{}
         \label{}
     \end{subfigure}
      \hspace{2px}
    \caption{{\footnotesize (a) Correlation plot between $ BR(\mu \rightarrow e + \gamma) $ and $ Sin^{2}\delta $ and (b) Correlation plot between $ BR(\mu \rightarrow e + \gamma) $ and $ \theta_{23} $  for allowed vacuum alignment (0,1,1) of triplet scalar field, $ \Phi_{s} $  involved in  both the inverse and linear seesaw models with normal hierarchy with  non-unitary parameter, $ \eta= 5.0 \times 10^{-7}$.}}
        \label{Fig011:4d}
\end{figure}

\begin{figure}[h]  
\centering
    \begin{subfigure}[b]{\textwidth}   
     \centering  
   \includegraphics[width=0.46\textwidth]{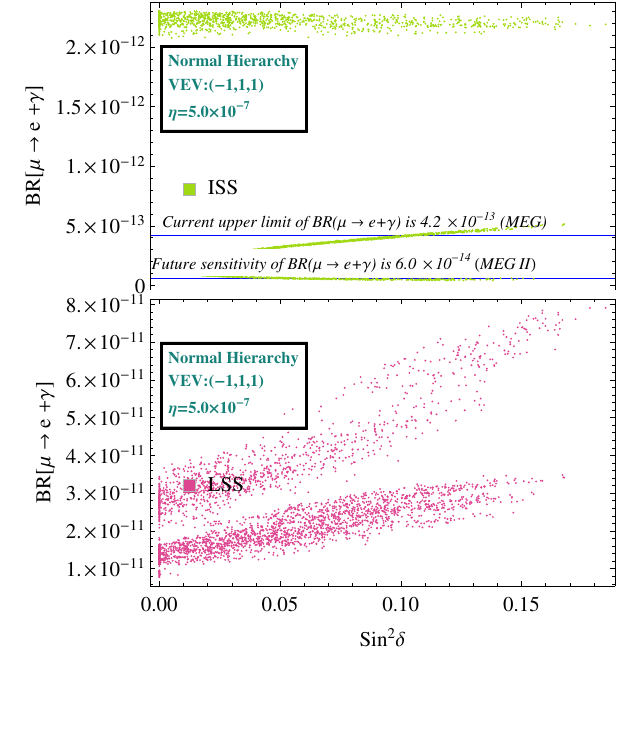} 
   \vspace{-0.001\baselineskip}
    \caption{}
       \hspace{10pt}
         \label{}
     \end{subfigure} 
  
    \begin{subfigure}[b]{\textwidth}
         \centering
   \includegraphics[width=0.46\textwidth]{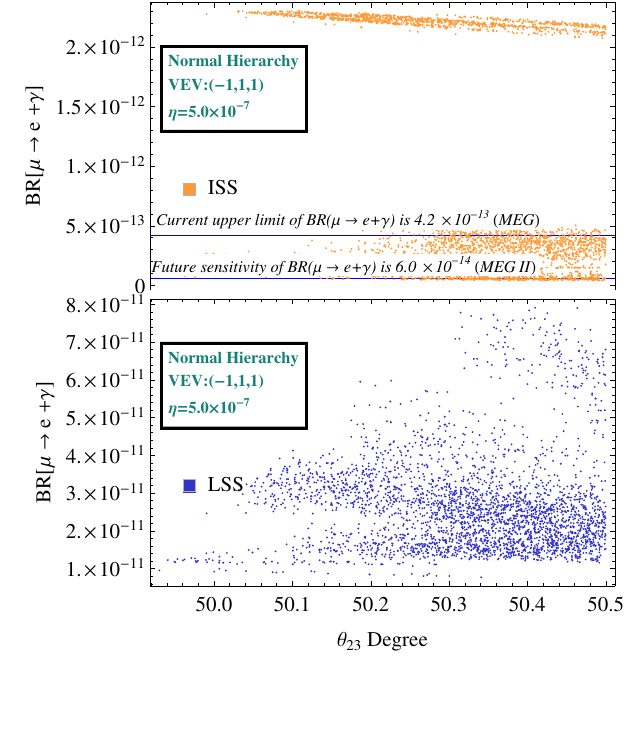}
  \vspace{-0.001\baselineskip}
    \caption{}
         \label{}
     \end{subfigure}
      \hspace{2px}
    \caption{{\footnotesize (a) Correlation plot between $ BR(\mu \rightarrow e + \gamma) $ and $ Sin^{2}\delta $ and (b) Correlation plot between $ BR(\mu \rightarrow e + \gamma) $ and $ \theta_{23} $  for allowed vacuum alignment (-1,1,1) of triplet scalar field, $ \Phi_{s} $  involved in  both the inverse and linear seesaw models with normal hierarchy with  non-unitary parameter, $ \eta= 5.0 \times 10^{-7}$.}}
        \label{Fig-111:4e}
\end{figure}
\section{Summary and Conclusions}
\label{sec: concl}
In this work, we  explored the feasibility  of cLFV $(\mu \rightarrow e + \gamma) $ decay  in ISS and LSS models with ${A}_{4} $ symmetry, and presented detailed analysis to pinpoint that which of the models could be more favourable for testing at $(\mu \rightarrow e + \gamma) $ experiments.  A few extra cyclic groups $ Z_4 $, $ Z_5$ and global symmetry $ U(1)_{x} $ were used to forbid contributions from unwanted terms to neutrino mass and ensuring only the contributions that generate the desired neutrino mass matrix for the two seesaw mechanisms. In our previous work \cite{Devi:2021ujp, Devi:2021aaz}, we found that only six VEV alignments of the triplet flavon $\Phi_s$ - (0,1,1)/(0,-1,-1) and (-1,1,1)/(1,-1,-1) for normal hierarchy and (0,-1,1)/(0,1,-1) for inverted hierarchy are allowed such that the unknown light neutrino oscillation parameters lie within their $ 3 \sigma $ range, for $A_{4}$ based  inverse and linear seesaw models (with a tolerance of $ < 10^{-5}$ for the all solutions). In this work we used these results and computed the branching ratio of $(\mu \rightarrow e + \gamma) $.  We did this  for seesaw scale $\sim 1$ TeV, $K_1= K_2=1$ eV,  $Y_{\mu_s} \sim 0.01$, $M_D \sim 10$ GeV, $\mu_s \sim 1$ keV, $M \sim 1$ TeV, $m_\nu\sim $ 0.1 eV, $\Lambda = 10^3 $ TeV, $M_L=1$ eV and $Y_D=0.05$.  We also chose some random values of the non-unitarity parameter $\eta$ within its currently allowed range.  Flavon VEVs, and (hence flavour symmetry breaking scale) show a rich variation of scales, as is seen from results in Table 5. We found that out of all cases in Table 4, for ISS, and for $\eta=5.0\times 10^{-7}$, VEV of $\Phi_s$(-1,1,1)/NH and (0,1,-1)/IH cases show closest agreement with the current bounds of MEG and sensitivity limits of MEG II experiment. So, the results of these cases validate our model. However, the BR for a lower value of  $\eta=9.0\times 10^{-9}$ will be testable at next generation experiments only, as their BR values lie beyond the sensitivity limits of planned experiment like MEG II. Other cases in Table 4. are ruled out as the BR projected by them is very high, and lies beyond the current limits of MEG experiment too. It  must be noted that decay rate of muon depends on the chosen value of the non-unitarity parameter $\eta$, flavon VEV alignment and MH of light neutrinos as well, hence, using the methodology of this work, and from future measurements at MEG II, it would be possible to pinpoint that which of the seesaw model, seesaw scale and VEV alignment of the triplet flavon is more favourable. Thus, the results of this work can throw light on dynamics of flavour symmetry as well as can help discriminate between LSS and ISS models, in context of cLFV decay ($\mu\rightarrow e\gamma$). Such computation can be done for other values of flavour symmetry breaking scale, couplings too and can be constrained with the current limits and sensitivity of cLFV experiments, which can tell about which of the two models/flavon VEV will be more favourable.

\section*{Acknowledgements}
Authors acknowledge support from FIST grant SR/FST/PSI-213/2016(C) dtd. 24/10/2017(Govt. of India) in upgrading the computer laboratory of the department where part of this work was done.

\end{document}